\documentclass[conference]{IEEEtran}
\IEEEoverridecommandlockouts 

\usepackage{cite} 
\usepackage{amsmath,amssymb,amsfonts} 
\usepackage{algorithmic} 
\usepackage{graphicx} 
\usepackage{textcomp} 
\usepackage{xcolor} 

\usepackage{url} 
\usepackage{booktabs} 
\usepackage{array} 

\usepackage{listings} 
\usepackage{enumitem} 
\usepackage{flushend} 

\def\BibTeX{{\rm B\kern-.05em{\sc i\kern-.025em b}\kern-.08em
    T\kern-.1667em\lower.7ex\hbox{E}\kern-.125emX}}

\lstdefinestyle{python}{
    language=Python,
    basicstyle=\ttfamily\footnotesize,
    keywordstyle=\color{blue}\bfseries,
    commentstyle=\color{green!50!black}\itshape,
    stringstyle=\color{red},
    showstringspaces=false,
    breaklines=true,
    frame=single,
    numbers=left,
    numberstyle=\tiny\color{gray},
    captionpos=b,
    columns=flexible,
    tabsize=4,
    xleftmargin=1em,
    framexleftmargin=0.5em,
}

\lstdefinestyle{semanticmessage}{
    basicstyle=\ttfamily\footnotesize,
    breaklines=true,
    frame=single,
    numbers=none, 
    xleftmargin=1em,
    framexleftmargin=0.5em,
    columns=flexible,
    captionpos=b,
    texcl=true, 
    escapeinside=`` 
}

\lstdefinestyle{appendixexample}{
    basicstyle=\ttfamily\scriptsize, 
    breaklines=true,
    frame=lines, 
    xleftmargin=0em,
    columns=flexible,
    showstringspaces=false,
    numbers=none, 
    rulecolor=\color{black!20}, 
    backgroundcolor=\color{black!5} 
}

\begin{document}

\title{Semantic-Aware Edge Intelligence for UAV Handover in 6G Networks}

\author{
    \IEEEauthorblockN{
        Aubida A. Al-Hameed\IEEEauthorrefmark{2},
        Mohammed M. H. Qazzaz\IEEEauthorrefmark{1}\IEEEauthorrefmark{2},
        Maryam Hafeez\IEEEauthorrefmark{1}, and
        Syed A. Zaidi\IEEEauthorrefmark{1}
    }
    \IEEEauthorblockA{
        \IEEEauthorrefmark{1}School of Electronic and Electrical Engineering, University of Leeds, Leeds, UK \\
        Email: \{ml14mmh, m.hafeez, s.a.zaidi\}@leeds.ac.uk
    }
    \IEEEauthorblockA{
        \IEEEauthorrefmark{2}College of Electronics Engineering, Ninevah University, Mosul, Iraq \\
        Email: aubida.alhameed@uoninevah.edu.iq
    }
}

\maketitle

\begin{abstract}
6G wireless networks aim to exploit semantic awareness to optimize radio resources. By optimizing the transmission through the lens of the desired goal, the energy consumption of transmissions can also be reduced, and the latency can be improved. To that end, this paper investigates a paradigm in which the capabilities of generative AI (GenAI) on the edge are harnessed for network optimization. In particular, we investigate an Unmanned Aerial Vehicle (UAV) handover framework that takes advantage of GenAI and semantic communication to maintain reliable connectivity. To that end, we propose a framework in which a lightweight MobileBERT language model, fine-tuned using Low-Rank Adaptation (LoRA), is deployed on the UAV. This model processes multi-attribute flight and radio measurements and performs multi-label classification to determine appropriate handover action. Concurrently, the model identifies an appropriate set of contextual "Reason Tags" that elucidate the decision's rationale. Our model, evaluated on a rule-based synthetic dataset of UAV handover scenarios, demonstrates the model's high efficacy in learning these rules, achieving high accuracy in predicting the primary handover decision. The model also shows strong performance in identifying supporting reasons, with an F1 micro-score of approximately 0.9 for reason tags. 
\end{abstract}


\section{Introduction}
\begin{figure}[!t]
    \centering
    \includegraphics[width=0.95\columnwidth]{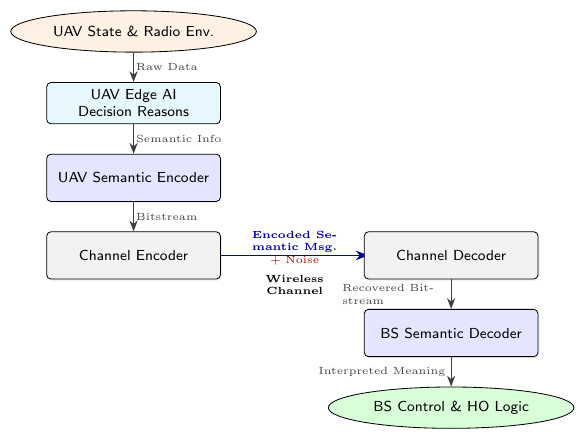} 
    \caption{Conceptual overview of the proposed semantic communication framework for UAV handover.}
    \label{fig:semantic_comm_overview}
\end{figure}

Recently, the expansion of Unmanned Aerial Vehicles (UAVs) in various applications has seen an unprecedented rise from surveillance to delivery, communication relaying, and intelligent network connectivity \cite{dai2022unmanned} . During a UAV mission, maintaining seamless connectivity is essential through handover mechanisms. Traditional handover decisions depend on the exchange of data measurements (such as the Received Power Reference Signal (RSRP), the Channel Quality Indicator (CQI)) between the UAV and the base station (BS). This results in considerable signaling overhead and potential delays \cite{khawaja2019uav}. The advent of 6G networks heralds a shift towards semantic communication \cite{gunduz2022beyond, yang2022semantic, kalita2024llms_iot}. In contrast to traditional communication systems that prioritize bit-level precision, semantic communication focuses on conveying the intended message (goal-related information), leading to more efficient and effective data transfer \cite{xie2021deep}. This approach is especially beneficial for resource-limited devices, such as UAVs, and in dynamic settings where understanding context is crucial. Edge intelligence facilitates processing near the data source, allowing localized decision making and decreasing dependence on centralized cloud processing \cite{shi2020edge, zhang2025large_aerial}. Figure \ref{fig:semantic_comm_overview} illustrates the proposed conceptual process of a semantic communication system for UAV handovers. Here, the UAV generates a succinct semantic message from the local data it processes. In this paper, we propose an edge-intelligent UAV handover system that embodies principles of semantic communication within a controlled environment. 
A lightweight language model, MobileBERT, fine-tuned using Low-Rank Adaptation (LoRA), is deployed on the UAV \cite{sun2020mobilebert} \cite{hu2021lora}. This model processes a textual representation of the complex radio and flight measurements to predict a primary handover decision from four distinct actions. Then identify a set of contextual "Reason Tags" that justify the decision based on predefined rules. After generating a concise semantic message for communication with the BS in service, the assessment and key rationale are provided. Our contributions include: (i) a multi-label classification framework for joint handover decision and reason tag prediction at the UAV edge (evaluated on a synthetic dataset); (ii) a demonstration of MobileBERT with LoRA for efficient fine-tuning on this task using synthetic data; and (iii) a system that generates semantic messages, discussing their potential for reducing communication overhead. Experimental results on a generated dataset of 5000 scenarios show near-perfect accuracy for the primary handover decision and strong performance in identifying relevant contextual reasons, highlighting the model's ability to learn the underlying data generation rules. This acts as an introductory validation, emphasizing the importance of further studies on applying this to real-world, non-rule-based contexts. The paper is structured as follows: Section \ref{sec:related_work} reviews related research. Section \ref{sec:system_model} details the design and approach of the system. Section \ref{sec:experimental_setup} describes the experimental setup, with results and analysis in Section \ref{sec:results}. Section \ref{sec:conclusion} concludes and suggests future work. Appendix \ref{app:semantic_examples} offers examples of generated semantic messages.
\section{Related Work}
\label{sec:related_work}
The semantic communication paradigm in the context of Shannon's information theory. This paradigm has gained attention with the emergence of 6G networks \cite{gunduz2022beyond}. This paradigm focuses on the transmission of meaningful information rather than the precise delivery of bits. Recent research explores the use of deep learning techniques for encoding and decoding semantic messages, highlighting significant compression efficiencies and robustness against channel noise by transmitting only the data relevant to the task \cite{xie2021deep, weng2021semantic}. Using Large Language Models (LLM) in semantic communication, especially within edge-based IoT networks, is an emerging investigative strand \cite{kalita2024llms_iot}. Our research aligns with this by concentrating on extracting and communicating a predefined "semantic essence" (decisions and justifications) from UAV handover assessments in a synthetic context. The management of UAV handover involves unique challenges due to aspects such as high mobility, three-dimensional motion, and strict Quality of Service (QoS) needs \cite{dai2022unmanned}. Traditional handover techniques based on fixed signal strength levels often fail. Hence, intelligent handover strategies powered by machine learning (ML) and artificial intelligence (AI) have been proposed \cite{mollel2021survey}. These include methods such as Q-learning, fuzzy logic, and neural network-based techniques to forecast handover requirements and select target cells \cite{amodu2025comprehensive} \cite{Qazzaz2024}. However, many current intelligent solutions still depend on sending vast amounts of raw data to centralized controllers or BSs for decision making. Enhancing network intelligence, particularly on the UAV itself, is vital for efficient and autonomous UAV functions \cite{shi2020edge, motlagh2017uav}. Edge AI can reduce latency, preserve backhaul capacity, and increase privacy. Onboard processing for UAV tasks such as navigation, object recognition, and network management is becoming more viable as lightweight deep learning models advance \cite{raza2025survey}. The convergence of large AI models with aerial edge devices, often involving edge-cloud collaboration, is a growing research domain that addresses the computational demands of these models \cite{zhang2025large_aerial}. Our research leverages the deployment of a fine-tuned MobileBERT model on UAVs to facilitate local handover assessments. Large Language Models (LLMs) and their scaled versions possess notable capabilities in comprehending contextual nuances and executing complex reasoning, even when confronted with text-based information \cite{devlin2018bert}. The integration of LLMs with UAVs represents a rapidly evolving area of study, aimed at exploring their ability to improve autonomy, perception, planning, and interaction between humans and UAVs \cite{tian2025uavs, javaid2024llms}. Fine-tuning models like BERT and MobileBERT for specific tasks is common \cite{sun2020mobilebert}. Effortless fine-tuning of LoRA parameters (PEFT) reduces trainable parameters, improving LLM for edge deployment \cite{hu2021lora}. Our method uses MobileBERT with LoRA to decode UAV scenario texts for multi-label classification, aligning sensor data with semantic labels through learned patterns.

\section{Proposed System Model and Methodology}
\label{sec:system_model}
Our proposed system empowers the UAV with edge intelligence to assess handover scenarios, make a primary decision, identify supporting reasons, and communicate this assessment semantically to the BS. The detailed architecture of the edge AI processing unit is depicted in Fig. \ref{fig:system_diagram_detail}. The core components are data representation, the fine-tuned language model, and semantic message generation. In this paper, "semantic understanding" refers to the model's learned capability to map the textualized input scenarios to the predefined set of decision and reason labels, based on the patterns inherent in the rule-based training data.

\subsection{Data Representation and Scenario Formulation}
The UAV continuously monitors its internal state (speed, buffer status, mission priority) and the radio environment (RSRP, RSRQ, CQI of the serving, target, and neighbor BS). To maximize the contextual understanding capabilities of a language model, these diverse parameters are formatted into a structured textual description for each handover assessment instance, as shown in Listing \ref{lst:scenario_text_example}. This textual representation serves as an input to our language model.
\begin{lstlisting}[style=python, caption=Example Formatted Scenario Text Input to Model, label=lst:scenario_text_example]
UAV Handover Assessment:
UAV State: Speed 15 m/s, Buffer 20%, Mission Priority High_Throughput.
Serving BS: ID BS3, RSRP -88.00 dBm, RSRQ -9.00 dB, CQI 10.
Handover Command: Handover to BS7.
Target BS (ID BS7): Local RSRP -105.00 dBm, Local RSRQ -14.00 dB, Local CQI 4.
Strongest Neighbor BS (ID BS4): Local RSRP -90.00 dBm, Local RSRQ -10.00 dB, Local CQI 9.
\end{lstlisting}
\subsection{Multi-Label Handover Assessment Model}
The essence of the edge intelligence system is a MobileBERT model, specifically fine-tuned for a multi-label head classification task through the use of LoRA, as depicted in Fig. \ref{fig:system_diagram_detail}. MobileBERT is selected due to its effective compromise between performance and computational efficiency, rendering it apt for on-device implementation \cite{sun2020mobilebert}.

\subsubsection{Label Schema}
The model selects labels from a list of 41 labels. These labels are organized into the following categories:

\begin{itemize}
    \item \textbf{Main Decision Classes:} These are exclusive labels indicating the primary handover action, as shown in Table \ref{tab:main_decision_classes_compact}.
    \item \textbf{Reason Tags (RT):} These are contextual labels that offer explanations or observations relevant to the scenario. These tags are subdivided into:
        \begin{itemize}
            \item \textit{Mutually Exclusive Groups:} In these groups, only one tag can be true for any given category. For example, in the context of Target RSRP Quality, relevant tags include \texttt{RT\_Target\_Excellent\_Signal\_RSRP} and \texttt{RT\_Target\_Good\_Signal\_RSRP}. In total, nine such groups account for variables such as target / current RSRP, CQI quality, RSRP advantage, UAV speed, buffer status, mission priority, and neighbor RSRP quality.
            \item \textit{Independent Tags:} Tags that can be true regardless of other tags being true or false, e.g., \texttt{RT\_Neighbor\_Is\_Stronger\_Alternative}, \texttt{RT\_Conflicting\_CQI\_RSRP\_Target}).
        \end{itemize}
\end{itemize}
Each training sample is tied to a multi-hot encoded vector, where $1$ signifies an active label, and $0$ otherwise. Our data creation protocol guarantees that each sample activates exactly one main decision label and selects one tag from each mutually exclusive reason tag group.

\subsubsection{Model Architecture and Fine-Tuning}
\begin{figure*}[!t]
    \centering
    \includegraphics[width=0.9\textwidth]{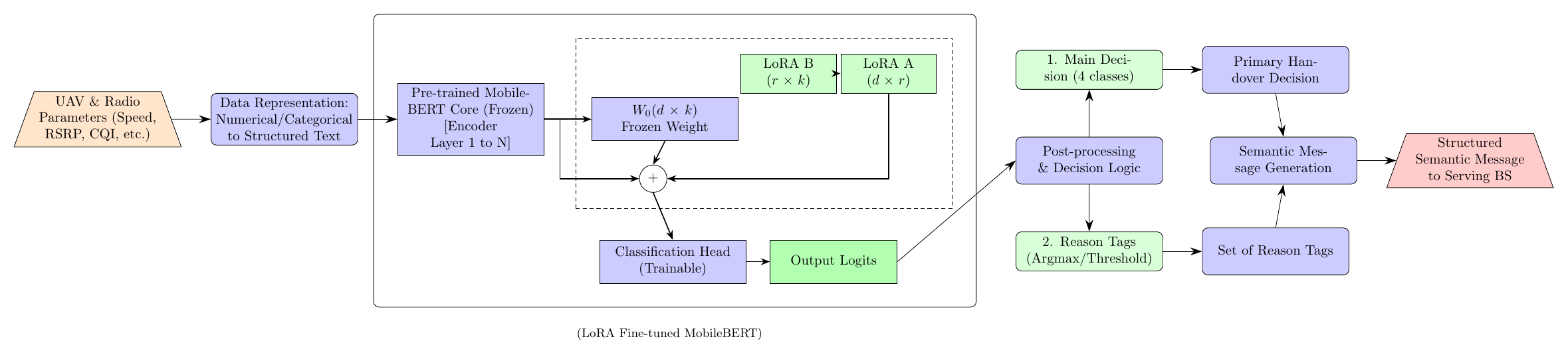} 
    \caption{Detailed system architecture of the proposed edge-intelligent UAV handover assessment framework. Raw UAV and radio parameters are converted to structured text, processed by a LoRA-fine-tuned MobileBERT model, and subjected to post-processing to yield a primary handover decision, reason tags, and a final semantic message.}
    \label{fig:system_diagram_detail} 
\end{figure*}

We use the pre-trained \texttt{ google / mobilebert uncased} model and adapt its classification head. Parameter-Efficient Fine-Tuning (PEFT) is used using LoRA \cite{hu2021lora}. LoRA injects trainable low-rank matrices (LoRA A and LoRA B in Fig. \ref{fig:system_diagram_detail} into the specified layers (query, key, value, attention output and dense layers of the feedforward network) of the frozen pre-trained model, significantly reducing the number of trainable parameters while maintaining strong performance. The model is trained using a binary cross-entropy with logit loss, appropriate for multi-label classification.

\begin{table}[t] 
\caption{Primary UAV Handover Decision Classes}
\centering
\footnotesize 
\renewcommand{\arraystretch}{0.9} 
\setlength{\tabcolsep}{3pt} 
\begin{tabular}{
    @{} 
    r
    @{\hspace{1ex}} 
    >{\raggedright\arraybackslash}p{2.2cm} 
    >{\raggedright\arraybackslash}p{%
        \dimexpr\columnwidth - 2.2cm - 2\tabcolsep - 1ex - 0.4cm - 2\arrayrulewidth\relax
    } 
    @{}
}
\toprule
\textbf{No.} & \textbf{Decision Class Label} & \textbf{Description} \\
\midrule
1. & \texttt{Exec\_HO\_Optimal} & Optimal to HO to target BS. \\
\addlinespace[0.3em]
2. & \texttt{Reject\_HO\_Curr\_Better} & Keep current BS; link superior. \\
\addlinespace[0.3em]
3. & \texttt{Reject\_HO\_Tgt\_Weak} & No HO; target signal insufficient. \\
\addlinespace[0.3em]
4. & \texttt{Question\_HO\_Conflict} & Review HO; data unclear/conflicting. \\
\bottomrule
\end{tabular}
\label{tab:main_decision_classes_compact}
\end{table}

\subsection{Post-processing and Semantic Message Generation}
The model produces raw logits for each of the 41 labels (see Fig. \ref{fig:system_diagram_detail}). A crucial post-processing step generates coherent, actionable data \begin{enumerate}
    \item \textbf{Main Decision Selection:} To determine the most likely primary handover decision, the \texttt{argmax} function is used on the logits associated with the initial \texttt{num\_main\_classes} (4) labels.
    \item \textbf{Reason Tag Selection:}
        \begin{itemize}
            \item Within each group of mutually exclusive reason tags, \texttt{argmax} is employed on their logits to choose the most likely tag from that group.
            \item For independent reason tags, probabilities calculated from the logits using a sigmoid function are compared to a preset threshold. In this study, a common threshold of 0.5 was utilized. Fine-tuning this threshold could be explored in future research.
        \end{itemize}
\end{enumerate}. The main decision, along with updated reason tags and key facts from the initial scenario, is synthesized into a structured semantic message (see Listing \ref{lst:semantic_message_example_results}). In the UAV Context section, "Interpreted as" expressions are derived from the model's predicted reason tags for each context parameter.

\section{Experimental Setup}
\label{sec:experimental_setup}
\subsection{Dataset Generation}
\label{subsec:dataset_generation}
Due to the lack of publicly available data sets tailored for this specific multi-label UAV handover assessment task, we generated a synthetic data set. The data set comprises 5000 unique UAV handover scenarios. Each scenario includes parameters such as UAV speed, buffer status, mission priority, and detailed radio measurements (RSRP, RSRQ, CQI) to serve, target, and BS nearest neighbor.

Data generation employs a rule-based system. This system assigns one of the four main decision classes and a set of relevant Reason Tags based on predefined conditions and thresholds applied to the scenario parameters. For instance, a \texttt{Reject\_Handover\_Target\_Signal\_Too\_Weak} decision might be triggered if the target BS RSRP falls below -100 dBm and its CQI is below 4. On the other hand, an \texttt{Execute\_Handover\_Optimal} decision might be favored if the target BS RSRP is significantly better (\textgreater 10 dB advantage) than the serving BS, and both target RSRP and CQI are above "Good" thresholds ( RSRP \textgreater -85 dBm, CQI \textgreater 9), and no conflicting UAV state parameters (like critical buffer for a high-throughput mission) are present. \texttt{Question\_Handover\_Conflicting\_Data} can arise if, for example, the target RSRP is high but CQI is low, or if the UAV mission priority conflicts with a high-speed UAV state when the signal quality difference between serving and target BS is marginal. The assignment of multiple reason tags aims to capture these interacting conditions. This rule-based approach was chosen to create a controlled environment that facilitates the initial validation of the model's ability. Also, learn complex input-output mappings. However, we acknowledge that this synthetic generation process, while incorporating parameter interactions, may not fully capture the stochasticity and unmodeled complexities of real-world environments. The data set was divided into 80\% for training (4000 samples) and 20\% for evaluation/test (1000 samples), stratified by the main decision class.

\subsection{Model and Training Parameters}
\begin{itemize}
  \item Our base model was \textbf{google/mobilebert-uncased}. We chose a LoRA configuration with Rank ($r$) = 16, Alpha ($\alpha$) = 32, and dropout = 0.1. These specific hyperparameters for LoRA were selected because they are commonly found in PEFT literature and provide a balance between the model's performance and the number of parameters that can be trained, which is ideal for initial assessments in resource-restricted edge environments. A detailed tuning of these hyperparameters was beyond the scope of our work. The targeted modules comprised the query, key, value, and attention output dense layers, along with the intermediate and output dense layers of the feedforward networks. For training, we used AdamW optimizer with a learning rate of $2 \times 10^{-5}$, a batch size of 16, over 40 epochs, incorporating a weight decay of 0.01, and a linear learning rate scheduler with a warm-up ratio of 0.1. The training was carried out on an NVIDIA GeForce RTX 2060 SUPER GPU.
\end{itemize}

\subsection{Evaluation Metrics}
Given the multi-label nature of the task, we use the following metrics \cite{Manning2008},\cite{Sokolova2009}:
\begin{itemize}
    \item \textbf{Main Decision Accuracy (Argmax):} Accuracy of predicting the correct single main decision class.
    \item \textbf{Main Decision F1 (Argmax):} Micro F1 score for main decision prediction.
    \item \textbf{F1-micro (Overall):} Micro-averaged F1-score across all labels.
    \item \textbf{F1-micro (Reason Tags):} Micro-averaged F1-score for the reason tags derived after postprocessing.
\end{itemize}

\section{Results and Discussion}
\label{sec:results}
\subsection{Quantitative Performance}
\begin{figure*}[!t]
    \centering
    \includegraphics[width=0.76\textwidth]{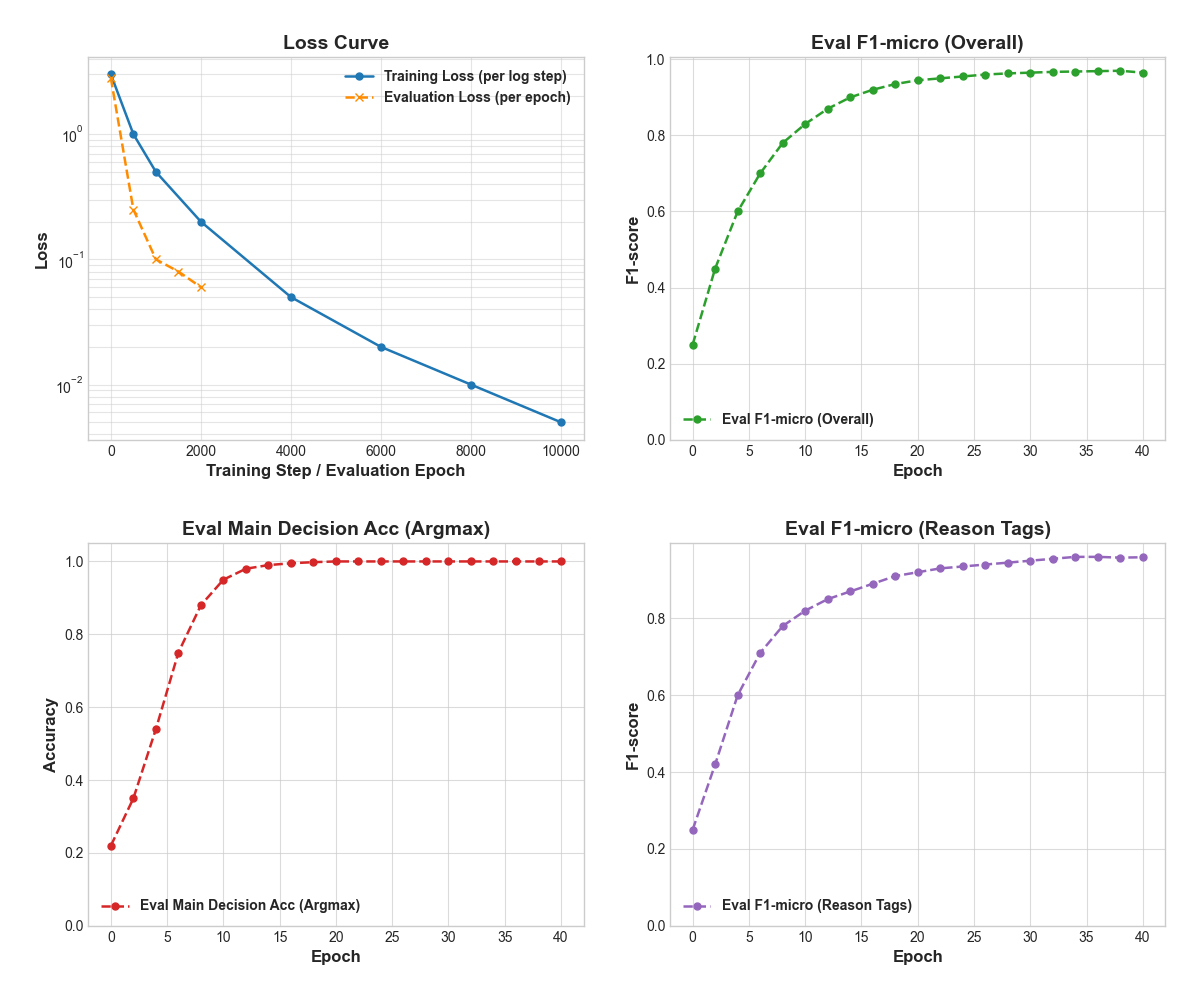} 
    \caption{Training dynamics observed on the validation set during training: (Top-Left) Loss Curve (log scale for y-axis), (Top-Right) Overall F1-micro. (Bottom-Left) Main Decision Accuracy (Argmax), (Bottom-Right) Reason Tags F1-micro.}
    \label{fig:training_plots} 
\end{figure*}

The model was trained for 40 epochs. The checkpoint demonstrating optimal performance during the validation phase was subsequently evaluated on a separate test set consisting of 1000 samples. Table \ref{tab:results} provides a summary of the principal performance metrics of the model evaluated on this set of tests. The model achieved an exceptional accuracy of 99.98\% in the prediction of the primary handover decision. The overall F1-micro score on all labels achieved a value of 0.9178, with a precision of 0.9193 and a recall of 0.9163. The F1-micro for processed reason tags was also very high at 0.9093. The loss curves of training and evaluation, along with key F1 scores and accuracies over epochs during the training phase, are presented in Fig. \ref{fig:training_plots}. The loss decreased consistently, indicating stable learning. The main decision accuracy (argmax) reached nearly perfect scores in training and maintained high performance. Also, the reason tag F1-micro scores showed improvement and demonstrated the model's effectiveness in learning the complex multilabel task.

\begin{table}[htbp]
\caption{Performance Metrics on the Test Set}
\begin{center}
\begin{tabular}{lc}
\toprule
\textbf{Metric} & \textbf{Value} \\
\midrule
Main Decision Accuracy (Argmax) & 0.9998 \\
Main Decision F1 (Argmax) & 0.9998 \\
\midrule
F1-micro (Overall) & 0.9178 \\
Precision-micro (Overall) & 0.9193 \\
\midrule
F1-micro (Reason Tags Only, Processed) & 0.9093 \\
Avg. Predicted Reason Tags & 9.60 \\
Avg. True Reason Tags & 9.63 \\
\bottomrule
\end{tabular}
\label{tab:results}
\end{center}
\end{table}

\begin{lstlisting}[style=semanticmessage, caption=Example Semantic Message Output (Reject Handover), label=lst:semantic_message_example_results]
UAV Assessment: Rejecting handover to Target BS BS7 due to weak signal. Key Factors: RSRP Relation: clear current advantage; Target Signal (RSRP poor, CQI low). UAV Context: Speed 15m/s (Interpreted as: medium speed uav), Buffer 20% (Interpreted as: critical low), Mission high throughput.
\end{lstlisting}

\subsection{Semantic Message Quality}
The system generates coherent semantic messages that reflect the model's assessment. An example semantic message for a scenario where the target BS was weak is shown in Listing \ref{lst:semantic_message_example_results}. In addition, additional examples illustrating different decision types are provided in the Appendix \ref{app:semantic_examples}. The qualitative interpretations ("medium speed uav," "critical low" buffer) in the "UAV context" part of these messages are directly derived from the model's predicted reason tags for those parameters.

It is important to note that occasionally misclassifications of specific reason tags can occur. For instance, as seen in Appendix A \ref{app:semantic_examples}, a scenario with a 95\% buffer was associated with the predicted "RT\_Buffer\_Critical\_Low" tag by the model. This led to a semantic interpretation where the 95\% buffer was incorrectly categorized as "critical low," a direct consequence of the model's output for that particular sample. These rare instances, despite high F1 scores, show that the model's interpretation of conflicting rules can sometimes differ from human intuition. This underscores the dual importance of the BS LLM's ability to comprehend potentially flawed UAV semantic messages, while also highlighting a key area for future investigation into the interplay between the two models' learning behaviors.

\subsection{Practical Deployment and Performance Profile}
\label{subsec:deployment_and_performance}
Deploying MobileBERT on resource-limited devices such as UAVs presents practical challenges. we performed a detailed resource analysis to quantify the footprint and readiness of our trained model.

\subsubsection{Hardware-Agnostic Indicators}
These metrics define the model's intrinsic complexity and storage requirements, independent of the execution hardware:
\begin{itemize}
    \item \textbf{Model Complexity:} The model comprises 27.13M total parameters, with the PEFT adapter accounting for only 0.02M trainable parameters.
    \item \textbf{Storage Footprint :} The trained model occupies 52.14 MB of disk space.
    \item \textbf{Computational Cost:} One inference requires 3.11 GFLOPs.
\end{itemize}

\subsubsection{Hardware-Specific GPU Benchmarks}
\begin{itemize}
    \item \textbf{Inference Latency:} An average latency of 66.72 ms per sample was achieved.
    \item \textbf{Throughput:} The system processes approximately 15 samples per second.
    \item \textbf{Memory Footprint:} Inference peak RAM usage was 970.75 MB.
\end{itemize}

\subsubsection{Deployment Considerations}
Real-time UAV handover decisions are critical and measured inference latency must be well within the network budget for the handover process. Furthermore, since our model was trained on synthetic data, ensuring its robustness and generalization to noisy, real-world sensor data is crucial. Future work should employ techniques such as resilient training or lifelong learning to bridge this gap.

\section{Conclusion and Future Work}
\label{sec:conclusion}
This paper presented an edge-intelligent UAV handover assessment system based on a LoRA-fine-tuned MobileBERT model, exploring its application within the semantic communication paradigm for future 6G networks using a rule-based synthetic environment. Evaluated on this data set of 5000 scenarios, our system achieves near-perfect accuracy (99.98\%) in predicting the primary handover decision on a held-out test set and provides a coherent set of supporting reason tags with a high F1 score ($\sim$0.91 for reason tags, $\sim$0.92 overall F1-micro). The generated semantic messages, when encoded, offer a pathway to significantly reduce communication overhead. This work successfully showcases how lightweight language models at the edge can learn to perform complex, context-aware decision-making based on patterns derived from structured textual input representing rule-based scenarios. While the performance on synthetic data is promising for validating the model's learning capacity, the critical next step is validation and adaptation using real-world UAV operational data to assess true generalization and robustness in environments not governed by predefined rules. Future work will focus on: (i) extensive testing with simulated datasets; (ii) developing adaptive or optimized thresholding mechanisms for independent reason tag selection; (iii) exploring more dynamic and context-sensitive semantic message generation techniques; (iv) Deploying the model as an O-RAN xApp in the BS for closed-loop semantic-driven handover automation; (v) Investigating alternative lightweight models and further quantization for enhanced on-device efficiency, considering the broader landscape of LLMs for UAVs, and (vi) conducting a rigorous analysis of error modes for reason tag predictions to further refine the model.

\appendix
\section{Illustrative Semantic Message Examples}
\label{app:semantic_examples}
This appendix provides four examples of semantic messages generated by the system for different handover scenarios from the test set, illustrating the various decision types. The "Interpreted as" phrases in the UAV Context part of the semantic message reflect the model's qualitative tag selection for those parameters, which are derived from the raw numerical inputs by extracting the string value from the predicted reason tag.

\subsection{Example A: Execute Handover Optimal}
\begin{lstlisting}[style=appendixexample, captionpos=none]
Raw Input Text:
UAV Handover Assessment:
UAV State: Speed 18 m/s, Buffer 95%, Mission Priority Low_Latency.
Serving BS: ID BS4, RSRP -109.81 dBm, RSRQ -19.40 dB, CQI 6.
Handover Command: Handover to BS1.
Target BS (ID BS1): Local RSRP -88.62 dBm, Local RSRQ -14.62 dB, Local CQI 12.
Strongest Neighbor BS (ID BS2): Local RSRP -71.52 dBm, Local RSRQ -6.47 dB, Local CQI 15.

Predicted Main Decision: Execute_Handover_Optimal
Predicted Reason Tags (Processed): ['RT_Target_Mediocre_Signal_RSRP', 'RT_Current_Poor_Signal_RSRP', 'RT_Target_CQI_High', 'RT_Current_CQI_Low', 'RT_Clear_Target_Advantage_RSRP', 'RT_Medium_Speed_UAV', 'RT_Buffer_Critical_Low', 'RT_Mission_Low_Latency', 'RT_Neighbor_Signal_Good', 'RT_Neighbor_Is_Stronger_Alternative']

Semantic Message to Serving BS:
UAV Assessment: Proposing handover to Target BS BS1. Key Factors: RSRP Relation: clear target advantage; Target Signal (RSRP mediocre, CQI high). UAV Context: Speed 18m/s (Interpreted as: medium speed uav), Buffer 95% (Interpreted as: critical low), Mission low latency.
\end{lstlisting}

\subsection{Example B: Question Handover Conflicting Data}
\begin{lstlisting}[style=appendixexample, captionpos=none]
Raw Input Text:
UAV Handover Assessment:
UAV State: Speed 30 m/s, Buffer 40%, Mission Priority Low_Latency.
Serving BS: ID BS10, RSRP -78.55 dBm, RSRQ -15.69 dB, CQI 11.
Handover Command: Handover to BS2.
Target BS (ID BS2): Local RSRP -79.73 dBm, Local RSRQ -19.40 dB, Local CQI 3.
Strongest Neighbor BS (ID BS9): Local RSRP -93.47 dBm, Local RSRQ -11.52 dB, Local CQI 8.

Predicted Main Decision: Question_Handover_Conflicting_Data
Predicted Reason Tags (Processed): ['RT_Target_Good_Signal_RSRP', 'RT_Current_Good_Signal_RSRP', 'RT_Target_CQI_Low', 'RT_Current_CQI_Medium', 'RT_Similar_RSRP', 'RT_High_Speed_UAV', 'RT_Buffer_High', 'RT_Mission_Low_Latency', 'RT_Neighbor_Signal_Mediocre', 'RT_Conflicting_CQI_RSRP_Target', 'RT_Unclear_Benefit_Due_To_Buffer_Mission']

Semantic Message to Serving BS:
UAV Assessment: Handover to Target BS BS2 requires review due to conflicting/unclear data. Key Factors: RSRP Relation: similar; Target Signal (RSRP good, CQI low); target RSRP/CQI conflicting; unclear benefit (buffer/mission constraint). UAV Context: Speed 30m/s (Interpreted as: high speed uav), Buffer 40% (Interpreted as: high), Mission low latency.
\end{lstlisting}
\subsection{Example C: Reject Handover Target Signal Too Weak}
\begin{lstlisting}[style=appendixexample, captionpos=none]
Raw Input Text:
UAV Handover Assessment:
UAV State: Speed 12 m/s, Buffer 25%, Mission Priority Standard.
Serving BS: ID BS7, RSRP -79.84 dBm, RSRQ -6.83 dB, CQI 9.
Handover Command: Handover to BS6.
Target BS (ID BS6): Local RSRP -115.79 dBm, Local RSRQ -14.08 dB, Local CQI 1.
Strongest Neighbor BS (ID BS10): Local RSRP -73.78 dBm, Local RSRQ -6.43 dB, Local CQI 14.

Predicted Main Decision: Reject_Handover_Target_Signal_Too_Weak
Predicted Reason Tags (Processed): ['RT_Target_VeryPoor_Signal_RSRP', 'RT_Current_Good_Signal_RSRP', 'RT_Target_CQI_Low', 'RT_Current_CQI_Medium', 'RT_Clear_Current_Advantage_RSRP', 'RT_Low_Speed_UAV', 'RT_Buffer_Sufficient', 'RT_Mission_Standard', 'RT_Neighbor_Signal_Good']

Semantic Message to Serving BS:
UAV Assessment: Rejecting handover to Target BS BS6 due to weak signal. Key Factors: RSRP Relation: clear current advantage; Target Signal (RSRP very poor, CQI low). UAV Context: Speed 12m/s (Interpreted as: low speed uav), Buffer 25% (Interpreted as: sufficient), Mission standard.
\end{lstlisting}

\subsection{Example D: Reject Handover Current BS Better}
\begin{lstlisting}[style=appendixexample, captionpos=none]
Raw Input Text:
UAV Handover Assessment:
UAV State: Speed 12 m/s, Buffer 70%, Mission Priority Standard.
Serving BS: ID BS6, RSRP -80.70 dBm, RSRQ -16.77 dB, CQI 12.
Handover Command: Handover to BS1.
Target BS (ID BS1): Local RSRP -100.82 dBm, Local RSRQ -7.07 dB, Local CQI 5.
Strongest Neighbor BS (ID BS7): Local RSRP -76.47 dBm, Local RSRQ -7.25 dB, Local CQI 15.

Predicted Main Decision: Reject_Handover_Current_BS_Better
Predicted Reason Tags (Processed): ['RT_Target_Poor_Signal_RSRP', 'RT_Current_Good_Signal_RSRP', 'RT_Target_CQI_Low', 'RT_Current_CQI_High', 'RT_Clear_Current_Advantage_RSRP', 'RT_Low_Speed_UAV', 'RT_Buffer_High', 'RT_Mission_Standard', 'RT_Neighbor_Signal_Good']

Semantic Message to Serving BS:
UAV Assessment: Maintaining current connection with BS BS6. Key Factors: RSRP Relation: clear current advantage; Target Signal (RSRP poor, CQI low). UAV Context: Speed 12m/s (Interpreted as: low speed uav), Buffer 70% (Interpreted as: high), Mission standard.
\end{lstlisting}

\end{document}